\documentclass[aps,prl,twocolumn,preprintnumbers,superscriptaddress,amsmath,amssymb]{revtex4}

\usepackage{graphicx}
\usepackage{subfigure}
\usepackage{epsfig}
\usepackage{xcolor}
\usepackage{dcolumn}
\usepackage{bm}
\usepackage{ulem}
\usepackage{color}
\usepackage{amsmath}
\usepackage{amsfonts}
\usepackage{amssymb}

\usepackage{braket}
\begin{document}
%\linenumbers

\title{Hidden charge density wave induced shadow bands and ultrafast dynamics of CuTe investigated using time-resolved ARPES}

\author{Haoyuan Zhong}
\affiliation
{State Key Laboratory of Low-Dimensional Quantum Physics and Department of Physics, Tsinghua University, Beijing 100084, P. R. China}

\author{Changhua Bao}
\affiliation
{State Key Laboratory of Low-Dimensional Quantum Physics and Department of Physics, Tsinghua University, Beijing 100084, P. R. China}

\author{Tianyun Lin}
\affiliation
{State Key Laboratory of Low-Dimensional Quantum Physics and Department of Physics, Tsinghua University, Beijing 100084, P. R. China}
 
\author{Fei Wang}
\affiliation
{State Key Laboratory of Low-Dimensional Quantum Physics and Department of Physics, Tsinghua University, Beijing 100084, P. R. China}

\author{Xuanxi Cai}
\affiliation
{State Key Laboratory of Low-Dimensional Quantum Physics and Department of Physics, Tsinghua University, Beijing 100084, P. R. China}

\author{Pu Yu}
\affiliation
{State Key Laboratory of Low-Dimensional Quantum Physics and Department of Physics, Tsinghua University, Beijing 100084, P. R. China}
\affiliation
{Frontier Science Center for Quantum Information, Beijing 100084, P. R. China}

\author{Shuyun Zhou}
\altaffiliation{Correspondence should be sent to syzhou@mail.tsinghua.edu.cn}
\affiliation
{State Key Laboratory of Low-Dimensional Quantum Physics and Department of Physics, Tsinghua University, Beijing 100084, P. R. China}
\affiliation
{Frontier Science Center for Quantum Information, Beijing 100084, P. R. China}

\date{\today}

\begin{abstract}
{\bf 
Revealing the fine electronic structure is critical for understanding the underlying physics of low-dimensional materials. Angle-resolved photoemission spectroscopy (ARPES) is a powerful experimental technique for mapping out the experimental electronic structure. By reducing the photon energy (e.g. to 6 eV) using laser sources, a greatly improved momentum resolution can be achieved, thereby providing opportunities for ``zooming in'' the fine electronic structure and even revealing the previously unresolvable bands near the Brillouin zone center. Here, by using quasi-one-dimensional material CuTe as an example, we demonstrate the unique capability of laser-based ARPES in revealing the fine electronic structures of ``hidden'' charge density wave induced shadow bands near the Brillouin zone center, which are previously unresolvable using synchrotron sources. The observation of the shadow bands reveals the CDW phase from the aspect of band folding, and the unpredicted CDW band hybridization strongly modifies the electronic structure and Fermi surface, which suggests that such hybridization must be taken into account for studying the CDW transition. Moreover, the ultrafast non-equilibrium carrier dynamics are captured by time-resolved ARPES, revealing the relaxation dynamics through electron-phonon scattering. Our work demonstrates the advantages of laser-based ARPES in zooming in the fine electronic structures, as well as capturing the ultrafast dynamics of low-dimensional materials.
}
\end{abstract}

\maketitle

The electronic structure E(k) determines the physical properties of low-dimensional materials, and therefore experimentally mapping out their electronic structure is critical. Angle-resolved photoemission spectroscopy (ARPES) is a powerful experimental technique for probing the electronic structure $E(k)$ of low-dimensional materials \cite{zxreve2003,zxrev2021,zhourev2022}. Figure 1(a) shows a schematic experimental setup for ARPES, where a probe light source excites electrons from the solid-state materials. The range of accessible in-plane momentum $k_{\parallel} = 0.512 \cdot\sin{\theta}\sqrt{h\nu -  E_B-\phi}$ is determined by the photon energy $h\nu$, where $\theta$ is the emission angle, $E_B$ is the binding energy of photoelectrons and $\phi$ is the work function. In order to cover a large energy and momentum range, synchrotron light sources with a photon energy of 20--100 eV are often used, which, however also inevitably leads to a compromised in-plane momentum resolution $\Delta k_{\parallel} = 0.512\cdot\cos{\theta}\sqrt{h\nu -  E_B-\phi}\cdot\Delta \theta$.
Reducing the photon energy from 100 eV to $\sim$ 6 eV using laser sources \cite{dessau_2006bscco, shinkbbf2008, xjzhoukbbf2008, xjzhoubscco, zhou2016mt, felix2016mt, felix2016wt, laserarpes_2018kondo, kondo_2021tase3, kondo_2020BSCCO} can significantly improve $\Delta k_{\parallel}$ by almost eight times (Fig.~1(b)), making low energy laser-ARPES  a potential ``microscope'' (see schematic illustration in Fig.~1(c)) for zooming in the fine electronic structures near the Brillouin zone (BZ) center, which are otherwise not resolvable using higher photon energy light sources generated from synchrotron. In addition, the ultrafast laser pulses also allow to capture the ultrafast carrier dynamics after exciting the sample with a pump beam by performing ARPES measurements in the pump-probe scheme, namely, time-resolved ARPES (TrARPES) \cite{TrARPES_rev_2024RMP,trarpes_2007bscco,trarpes_2008tbte3,trarpes_2012bs,trarpes_2012tise2,TrARPES_FeSe_ZX2017,trarpes_2019bscco,bao2022light,TrARPES_Ulstrup}.

Here, we demonstrate the advantages of laser-ARPES in revealing the fine electronic structures near the BZ center of various quantum materials, such as Li-intercalated graphene, topological semimetal PtSn$_4$, charge density wave (CDW) materials IrTe$_2$ and CuTe. Using CuTe as an example \cite{thomas2013cute,zsy2018cute,kwon2024dual}, we further show that the ``hidden" CDW shadow bands, which were unresolvable in previous ARPES measurements using synchrotron light sources  \cite{zsy2018cute}, can now be clearly resolved using $\sim$6 eV laser sources, and the CDW-induced band hybridization is identified below the CDW transition temperature. Moreover, by performing TrARPES with a pump pulse, the relaxation dynamics of photo-excited carriers are captured, revealing the role of electron-phonon (el-ph) scattering the non-equilibrium state. Our work provides new insights into the CDW physics of CuTe, and demonstrates the advantages of low energy laser-ARPES in zooming in the fine electronic structures near the BZ center as well as revealing the related non-equilibrium carrier dynamics.

\begin{figure*}[htbp]
		\centering
		\includegraphics[width=16.9 cm] {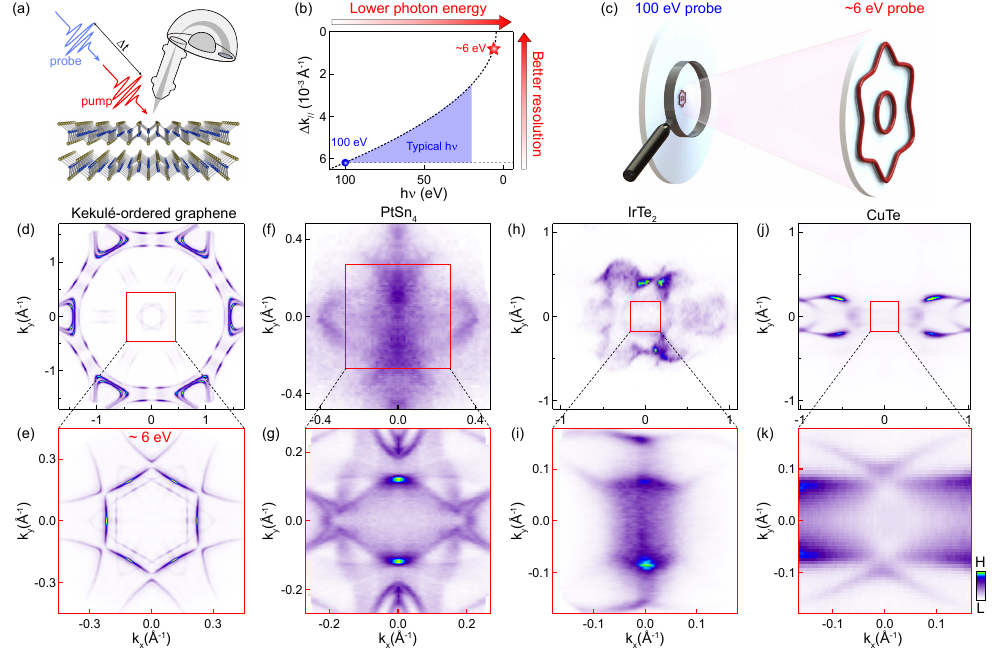}
		\label{Fig1}
		\caption{Zooming in the fine electronic structures by laser-ARPES. (a) Schematics of laser-based TrARPES setup. (b) The momentum resolution as a function of photon energy. The work function $\phi$ is 4.5 eV, $\theta$ is 45$^{\circ}$, $\Delta \theta$ is 0.1$^{\circ}$ in calculation. (c) Schematics of the ``microscope'' ability of zooming in Fermi surface structures of laser-ARPES. (d) and (e) Fermi surface of Kekul\'e-ordered graphene measured by (d) helium lamp source at $h\nu$ = 21.2 eV and (e) laser source with $h\nu$ = 6.2 eV at 80 K.  (f) and (g) Fermi surface map of PtSn$_4$ measured by (f) helium lamp source with $h\nu$ = 21.2 eV and (g) laser source (symmetrized with respect to k$_x$ and k$_y$ = 0) with $h\nu$ = 6.7 eV at 80 K. (h) and (i) Fermi surface map of IrTe$_2$ measured by (h) synchrotron source  with $h\nu$ = 100 eV and (i) laser source with $h\nu$ = 6.2 eV at 80 K  \cite{szhouirte2022cp}. (j) and (k) Fermi surface map of CuTe measured by (j) synchrotron source with $h\nu$ = 80 eV at 20 K  \cite{zsy2018cute} and (k) laser source with $h\nu$ = 6.3 eV at 80 K. }\label{Fig1}
	\end{figure*}

\begin{figure*}[htbp]
		\centering
		\includegraphics[width=16.9 cm] {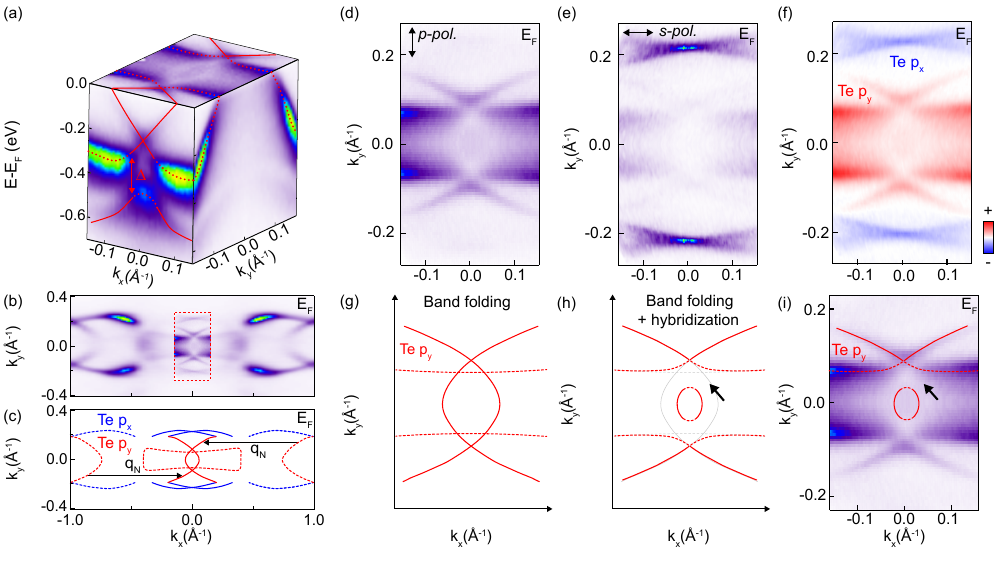}
		\label{Fig2}
		\caption{Revealing the ``hidden'' CDW shadow bands by laser-ARPES. (a) Three-dimensional electronic structure of CuTe by laser-ARPES. (b) Fermi surface map of CuTe measured using high photon energy ($h\nu$ = 80 eV). The inset is sum of the laser-ARPES data measured at $h\nu$ = 6.3 eV with $s$ and $p$ light polarizations. (c) Schematics of Fermi surface of CuTe due to CDW-induced band folding. Red and blue curves represent Te $p_y$ and $p_x$ contribution  \cite{zsy2018cute}. Dashed curves are the original pockets while solid curves are replicas translated by the CDW vectors. (d) and (e) Fermi surface maps using $p-pol.$ and $s-pol.$ laser polarizations respectively (symmetrized with respect to k$_y$ = 0), and the directions of the electrical field are marked by the arrows. (f) Differential intensity by subtracting (e) from (d). (g) and (h) Schematic of  replica pockets without (g) and with (h) band hybridization.   The band hybridization between the original pockets and folded pockets opens up a gap at the crossing points. (i) Fermi surface map with schematic in (h) over-plotted.}\label{Fig2}
		
	\end{figure*}
	
Low-energy laser-ARPES can not only ``magnify'' the electronic structures near the BZ center, more interestingly it can reveal the previously unresolvable ``hidden'' bands \cite{xjzhoubscco, kondo_2020BSCCO} as demonstrated in Figs.~1(d)--(k). Figure 1(d) shows the Fermi surface map of Li-intercalated graphene where the pockets near the K points are folded to the $\Gamma$ point by the ($\sqrt{3}\times\sqrt{3}$)R30$^\circ$ Kekul\'e order \cite{szhoukekule2021prl, szhoukekule2022prb}. While the three pockets corresponding to trilayer graphene are clearly observed near the BZ corners using helium lamp source, only one set of replica pockets is observed near the $\Gamma$ point \cite{szhoukekule2021prl,szhoukekule2022nsr}. 
Laser-ARPES measurements instead can zoom in the fine electronic structures near the BZ center, and resolve the other two folded pockets as shown in Fig.~1(e). In topological semimetal PtSn$_4$ which has a complex Fermi surface structure \cite{ptsn4_2012prb_canfield, ptsn4_2016np_kaminski}, only some broad features are observed in the Fermi surface map in Fig.~1(f) using a helium lamp source. Interestingly, laser-ARPES measurements successfully resolve a nice and complex fine electronic structure in Fig.~1(g). For CDW materials, the shadow bands induced by band folding due to the CDW periodicity are typically much weaker than the original bands, which are therefore often ``hidden'' or unresolvable as compared to the strong original bands. Laser-ARPES measurements provide new opportunities for resolving such ``hidden'' shadow bands. Figures~1(h)--(k) show comparisons between the Fermi surface maps measured with high and low photon energies in two CDW materials IrTe$_2$ \cite{IT_2014NJP_DingH, IT_2015PRL_Kyoo, szhouirte2022cp} and CuTe \cite{zsy2018cute}, where CDW-induced shadow bands are clearly resolved as shown in Figs.~1(i) and 1(k), demonstrating the power of laser-ARPES measurements in magnifying the ``hidden'' shadow bands. Moreover, the probe laser source with tunable photon energy provides opportunities for overcoming the dipole element effects, which either enhance or suppress the intensity of the bands (see Fig.~S1 in Supplemental Material \cite{supp}). The four types of materials presented above demonstrate that laser-ARPES measurements can provide opportunities for zooming in the fine electronic structures, and even revealing the previously unresolvable bands of quantum materials near the $\Gamma$ point. Especially for CuTe, revealing the ``hidden'' shadow bands helps to understand the CDW state from another aspect (folded bands due to new lattice periodicity), in complementary to the aspect of CDW energy gap \cite{zsy2018cute}. Therefore, below we further use CuTe as an example to explore the physics of the ``hidden'' bands, including evidence of band folding and CDW-induced hybridization gap as well as the ultrafast dynamics.

\begin{figure*}[htbp]
		\centering
		\includegraphics[width=16.9 cm] {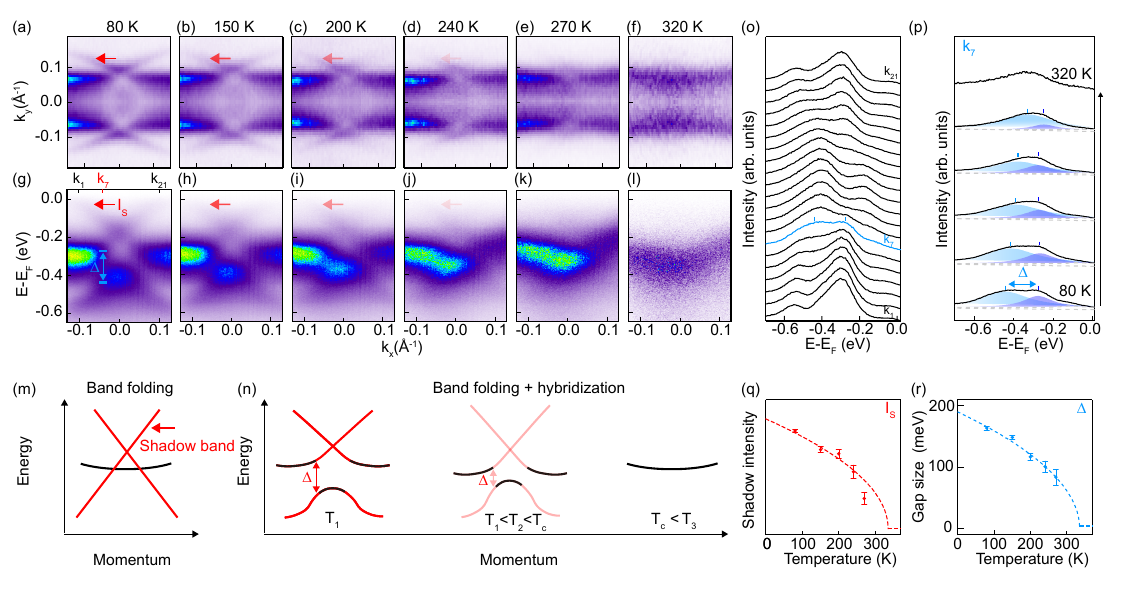}
		\caption{Temperature dependence of CDW folding and hybridization. (a--f) Fermi surface maps with temperature ranging from 80 K to 320 K (symmetrized with respect to k$_y$ = 0). (g--l) Corresponding dispersion images along $k_y$ = 0.12\AA $^{-1}$ in (a), where the $I_S$ is the intensity of the shadow bands and $\Delta$ is the hybridization gap. (m) Schematic of the electronic structure without hybridization. (n) Schematic summary of CDW melting when increasing the temperature: decreasing of hybridization gap and weakening of shadow band. (o) Extracted EDCs from $k_1$ to $k_{21}$ marked in (g). (p) Extracted EDCs at momentum position $k_7$ in (o) from 80 K to 320 K, which are fitted by three Lorentzian peaks and the main peaks are filled blue.  (q) Extracted shadow band intensity $I_S$ (see Fig.~S2 for raw data in Supplemental Material \cite{supp}), the red dashed line is BCS-type gap equation with transition temperature $T_c$ = 335 K. (r) Extracted gap size from (p) as function of temperature, the blue dashed line is BCS-type gap equation with transition temperature $T_c$ = 335 K. We note that the gap is hardly distinguished at 320 K due to the broad bandwidth.}\label{Fig3}
	\end{figure*}

CuTe is a quasi-one-dimensional room-temperature CDW material \cite{thomas2013cute}, whose CDW band structure has been investigated by synchrotron-based ARPES measurements \cite{zsy2018cute}, while CDW-induced shadow bands remain elusive. The successful observation of CDW shadow bands by laser-ARPES allows to investigate the CDW phase transition of CuTe from a new perspective, namely, the evolution of the shadow bands. Figure~2(a) shows the full three-dimensional electronic structure of CuTe using laser-ARPES. To further confirm that the observed new bands near the $\Gamma$ point are from CDW shadow bands, we compare the fine Fermi surface map (marked by red dashed box) with a large-ranged Fermi surface map measured by a high photon energy in Fig.~2(b). As schematically summarized in Fig.~2(c), the CDW nesting vectors $q_x = \pm 0.4a^{*}$ \cite{zsy2018cute} translates the pockets (dashed curves) originated from Te $p_x$ (blue) and $p_y$ (red) orbitals near the BZ boundary to the $\Gamma$ point, and the replica pockets (solid curves) match well with the measured Fermi surface by laser-ARPES, thereby suggesting that they are likely shadow bands induced by the CDW order.

Polarization-dependent laser-ARPES measurements are performed to further confirm the symmetry properties of these shadow bands. Figures 2(d) and 2(e) show a comparison of the Fermi surface maps using two different light polarizations:  $p-pol.$, where the electric field is along $k_y$ direction; and $s-pol.$, where the electric field is along $k_x$ direction.  Distinct intensity distributions are observed, and the differential (dichroic) intensity spectrum is shown in Fig.~2(f). The polarization-dependent laser-ARPES measurements show that the pockets around  $\Gamma$ (red) and away from $\Gamma$ (blue) are from Te $p_y$ and $p_x$ respectively (similar to the polarization-dependent ARPES measurements in another $p$ orbital system \cite{PolARPES_2015PRB}), which are also consistent with the symmetry properties of the original bands \cite{zsy2018cute}. However, the simple band folding picture in Fig.~2(g) does not fully match the measured Fermi surface map shown in Fig.~2(i), suggesting there is significant hybridization between the shadow bands (red solid curves) and original bands (red dashed curves), as schematically illustrated in Fig.~2(h). Such hybridization also results in the gap opening  in Fig.~2(a).

The CDW phase transition can be directly visualized by tracing the evolution of the shadow bands and the hybridization gap with temperature. As shown in Figs.~3(a)--(f), the shadow bands marked by red arrows gradually disappear with increasing temperature, and eventually become undetectable at 320 K, which is near the CDW phase transition temperature. Similar temperature evolution is also observed in the dispersion images shown in Figs.~3(g)--(l). Moreover, the hybridization gap (indicated by the blue arrow in Fig.~3(g)) also decreases gradually with increasing temperature. The electronic structure without hybridization is schematically shown in Fig.~3(m), where hybridization between shadow and origin bands opens up a hybridization gap. With increasing temperature, both the shadow band intensity and the hybridization gap decrease when approaching the CDW phase transition as shown in Fig.~3(n). A quantitative analysis of the shadow band intensity is shown in Supplemental Material Fig.~S2 \cite{supp}
, and analysis of the hybridization gap is performed by fitting the energy distribution curves (EDCs) measured at the momentum $k_7$ (the blue curve in Fig.~3(o)) with different temperatures as shown in Fig.~3(p). The evolution of the shadow band intensity (Fig.~3(q)) and the hybridization gap (Fig.~3(r)) is consistent with BCS-type gap equation \cite{chen2015charge, zsy2018cute, VT2019PRB_SZ}, suggesting that the weakening of shadow bands and the decreasing of the hybridization gap are both induced by the suppression of the CDW order. The unpredicted band hybridization between the shadow bands and original bands not only opens hybridization gaps but also strongly modifies the ``topology'' of Fermi surface which should have impact on its physical properties (e.g. transport properties). Such observation demonstrates the important role of band hybridization between shadow bands and original bands in CDW transition, which has been however largely ignored in previous studies. Our new finding suggests that such hybridization must be taken into account for studying the CDW transition in the future. The observation of CDW-induced shadow bands and hybridization gap, made possible by low-energy laser-ARPES measurements, provides new perspectives for the electronic structure modification of CuTe in the CDW phase.

\begin{figure*}[htbp]
\centering
\includegraphics[width=16.9 cm] {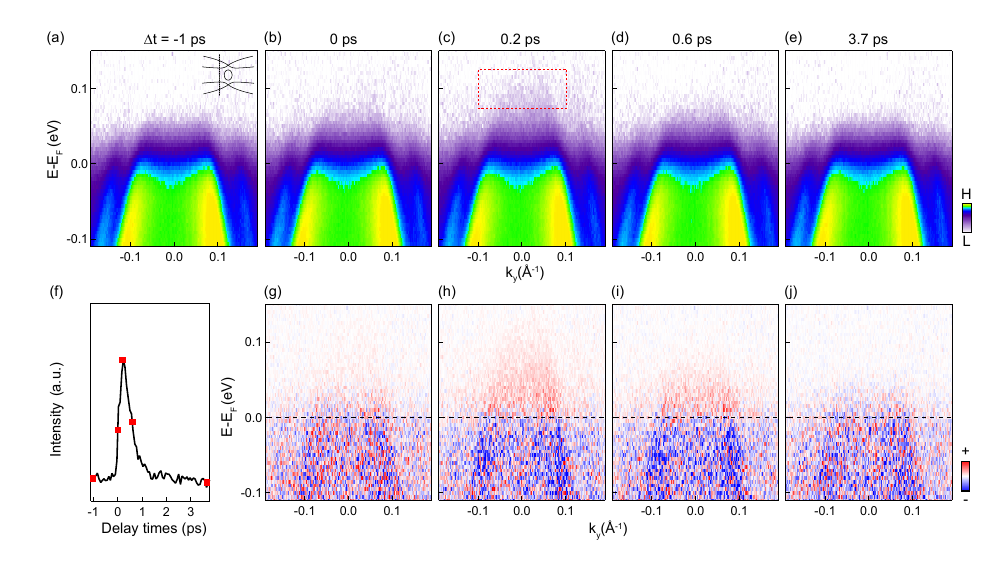}
\caption{Transient electronic structure of CuTe upon photoexcitation. (a--e) Snapshots of electronic dispersion (dispersion along $k_x$ = -0.08 \AA$^{-1}$ illustrated in the inset) at different pump-probe delay times marked in (f). (f) Temporal evolution of photoexcited electrons integrated by the red dashed box in (c). (g--j) Differential TrARPES spectra at different delay times after subtracting the spectrum at -1 ps. The pump and probe photon energies are 1.58 eV and 6.3 eV, respectively, with a pump fluence of 280 $\mu$J/cm$^2$ measured at 80 K.}\label{Fig4}
\end{figure*}

The ultrafast probe laser pulses have another advantage: the capability to measure the ultrafast carrier dynamics by combining a pump beam in TrARPES measurements. Meanwhile, revealing the ultrafast dynamics is important for CDW materials  \cite{trarpes_2008tbte3,trarpes_2012tise2,trarpes_2019tise2,trarpes_2020tase2}, because electron-phonon (el-ph) and electron-electron (el-el) scattering can drive the relaxation dynamics. We first investigate the electronic structure evolution of CuTe upon photo-excitation.  Figures~4(a)--(e) show TrARPES snapshots of CuTe at different delay times, where the representative delay times are marked in the time trace (integrating red box in Fig.~4(c)) in Fig.~4(f). The photo-excited electronic states can be better revealed in the differential images in Fig.~4(g)--(j), which are obtained by subtracting the dispersion image by data measured at -1 ps. A fast relaxation dynamics is clearly distinguished by comparing data at 0.2 ps (Fig.~4(h)) and 0.6 ps ((Fig.~4(i)), which is quantitatively analyzed in Fig.~5.

\begin{figure*}[htbp]
\centering
\includegraphics[width=16.9 cm] {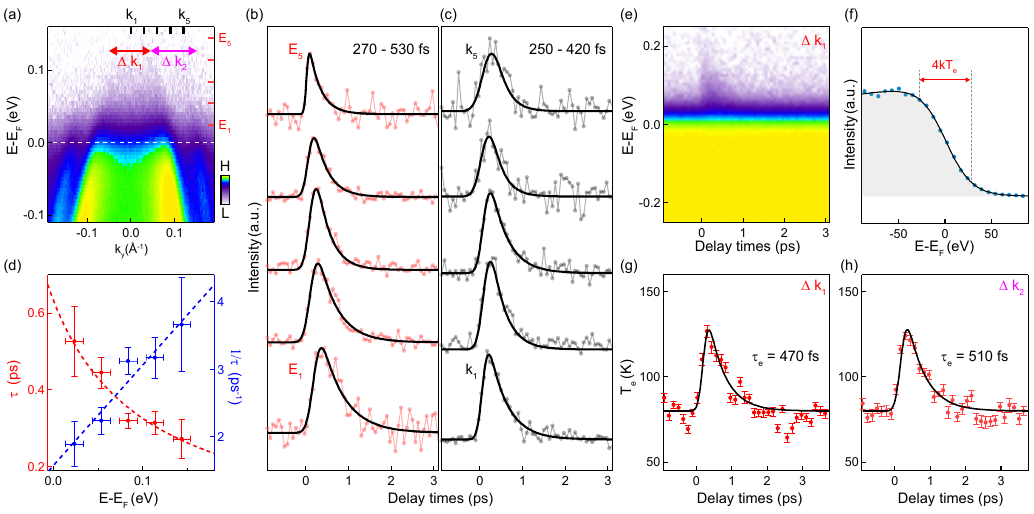}
\caption{Ultrafast relaxation dynamics of CuTe upon photo-excitation. (a) Snapshot of electronic structure (along $k_x$ = -0.08 \AA$^{-1}$) at pump-probe delay time $\Delta t$ = 0.2 ps. (b) Temporal evolution of photo-excited electrons at a few selected energies (integrate momentum range of -0.05--0.05 \AA) indicated by red tick marks E$_1$--E$_5$ in (a). (c) Temporal evolution of photo-excited electrons at a few selected momentums (integrate energy range of 0.05--0.15 eV) indicated by black tick marks k$_1$--k$_5$ in (a).  (d) Extracted lifetime $\tau$ (red) and scattering rate $1/\tau$ (blue) of photo-excited electrons as a function of energy, where $1/\tau$ is fitted by a line. (e) Integrated EDCs at momentum range $\Delta k_1$ marked in (a) as a function of delay time. (f) EDC at delay time of -500 fs from (e) and corresponding Fermi-Dirac fitting. (g, h) Extracted electronic temperature as a function of delay time (g) near $\Gamma$ point ($\Delta k_1$ in(a)) and (h) away from $\Gamma$ point ($\Delta k_2$ in(a)). Note that the electronic temperature increase in CuTe is low, similar to the case in metal \cite{Te_copper,Te_gold,Te_VTe2}, which is due to high electron density of state near Fermi level. The pump and probe photon energies are 1.58 eV and 6.3 eV, respectively, with a pump fluence of 280 $\mu$J/cm$^2$ measured at 80 K. Time resolution and energy resolution are shown in Fig.~S3 in Supplemental Material \cite{supp}. }\label{Fig5}
\end{figure*}

The energy and momentum resolved carrier dynamics is further revealed, which helps us to distinguish the relaxation dynamics. The energy and momentum dependent temporal evolution is shown in Fig.~5(b) and 5(c), which is extracted from energies and momentums marked in Fig.~5(a). For all energies and momentums, the relaxation dynamics lies between 200--600 fs, which are obtained by fitting the temporal evolution of the ARPES intensity  \cite{TrARPES_RSI2021}. Here, we further analyze the energy-dependent lifetime $\tau$ (from Fig.~5(b)) and scattering rate $1/\tau$, which are plotted in Fig.~5(d). The lifetime is consistent with the recent complimentary reflectivity measurements of the same material \cite{wnl2022cute}. Such relaxation time is much faster compared to many quantum materials, for example, Bi$_2$Se$_3$ \cite{trarpes_2012bs,gedik2012bs} and Cd$_3$As$_2$ \cite{szhou2022ca}, which might  indicate strong el-ph coupling (inversely proportional to relaxation times \cite{EPC_metal_1987PRL, szhou2022ca}) in CuTe, as theoretical calculations \cite{zsy2018cute, cute_epc2019, cute_epc2021, cute_epc2023_PRB} and transport measurement \cite{cute_transport2020} indicated. We note that the el-ph coupling induced kink feature is not resolvable in our measurements, which is not surprising considering that the phonon energy is quite low, $\sim 10$ meV,  smaller than the energy resolution of  $46$ meV (see Fig.~S3 for energy resolution in Supplemental Material \cite{supp}). In addition to the carrier dynamics, the photo-excited thermodynamics can also be revealed. Figure~5(e) shows the continuous evolution of EDCs as a function of delay times by integrating momentum range $\Delta k_1$ in Fig.~5(a), where the transient electronic temperature is extracted by fitting the EDCs with a Fermi-Dirac distribution, as shown in Fig.~5(f). The evolution of the electronic temperature $T_e$ shows a relaxation time of 470 $\pm$ 70 fs as shown in Fig.~5(g) near $\Gamma$ point ($\Delta k_1$ in Fig.~5(a)) and a relaxation time of 510 $\pm$ 60 fs as shown in Fig.~5(h) away from $\Gamma$ point ($\Delta k_2$ in Fig.~5(a)), which are both $\sim$ 500 fs. First, compared to the characteristic time scale, which is 10--100 fs for el-el scattering and is 100--1000 fs for el-ph scattering \cite{trarpes_2012tise2}, we can imply that photo-carriers of CuTe relax through el-ph scattering. Second, relaxation time scale of CuTe is comparable to many CDW materials, for example, TbTe$_3$ \cite{trarpes_2008tbte3}, 1T-TaS$_2$  \cite{TaS2_2006PRL,trapres_2023tas2}, 1T-TiSe$_2$ \cite{trarpes_2019tise2} and 1T-TaSe$_2$ \cite{trarpes_2020tase2} (all of them relax with time scale of few hundreds of femtoseconds). While those CDW materials relax through el-ph scattering, we indicate that the relaxation dynamics of CuTe is driven by el-ph scattering.  We would like to note that el-el scattering (supported by calculation \cite{zsy2018cute}) cannot be fully ruled out since the time scale of tens of femtoseconds might not be distinguished.

In summary, by using laser-ARPES with low photon energy, a significantly improved momentum resolution is achieved, allowing to successfully reveal the  fine electronic structures near the $\Gamma$ point for various  quantum materials. Using CuTe as an example, the CDW phase transition is revealed from the view of  ``hidden'' shadow bands. Interestingly, the unpredicted CDW band hybridization strongly modifies the electronic structure and Fermi surface, which suggests that such hybridization must be taken into account for the physics of the CDW transition. Moreover, the ultrafast non-equilibrium carrier dynamics are captured by time-resolved ARPES, revealing the relaxation dynamics through el-ph  scattering. Our work demonstrates the power of low energy laser-ARPES in zooming in the fine electronic structures and ultrafast dynamics near the BZ center, which can be extended to a wide range of quantum materials.

\begin{acknowledgments}
\section*{ACKNOWLEDGMENTS}
This work is supported by the National Key R$\&$D Program of China (No.~2021YFE0107900, 2020YFA0308800, 2021YFA1400100), the National Natural Science Foundation of China (Grant No.~12234011, 92250305, 52388201, 11725418  and 11427903). C.B. is supported by a project funded by China Postdoctoral Science Foundation (No.~2022M721886 and No.~BX20230187) and the Shuimu Tsinghua Scholar Program.
\end{acknowledgments}

\subsection{Conflict of interest}
The authors have no conflicts to disclose.
%\bibliography{ctreference}

\end{document}